\documentclass[aps,showpacs,twocolumn,floats]{revtex4}
\usepackage{graphicx}
\usepackage{amsmath}
\usepackage{amssymb}
\usepackage{psfrag}

\begin{document}

\newcommand{\ee}[1]{\label{#1} \end{equation}}
\newcommand{\be}{\begin{equation}}
\newcommand{\av}[1]{\left\langle #1 \right\rangle }
\def\reff#1{(\ref{#1})}
\def\re{\mbox{Re}}
\def\w{\omega}
\def\W{\Omega}
\def\t{\tau}
\def\e{\varepsilon}

\title{
 Partially integrable dynamics
of hierarchical populations of coupled oscillators}
\author{Arkady Pikovsky and Michael Rosenblum}
\affiliation{Department of Physics and Astronomy, Potsdam University,
Karl-Liebknecht-Str 24, D-14476 Potsdam-Golm, Germany}
\date{\today}
\begin{abstract}%
We consider oscillator ensembles consisting of subpopulations of identical units, 
with a general heterogeneous coupling between subpopulations. 
Using the Watanabe-Strogatz ansatz we 
reduce the dynamics of the ensemble to a relatively small number of 
dynamical variables plus constants of motion. This reduction is
independent of the sizes of subpopulations and remains valid in the
thermodynamic limits. The theory is applied to the standard Kuramoto
model and to the description of two interacting subpopulations, where we report 
a novel, quasiperiodic chimera state.
\end{abstract}

\maketitle

Large populations of coupled oscillators occur in a variety of applications 
and models of natural phenomena, ranging from collective dynamics 
of multimode lasers and Josephson junction arrays to the pedestrian synchrony 
\cite{Kuramoto-84};
the analysis of the dynamics of these systems is a topic of high interest.
Even in the context of the simplest,  paradigmatic case of 
globally (all-to-all) connected, sine-coupled phase oscillators 
(the famous Kuramoto model and its generalizations), many problems remain yet 
unsolved, especially
those related to a heterogeneous coupling and nontrivial collective dynamics.
In this Letter we treat an important case of a {\em hierarchically}
 organized
population. 
It can be viewed as a (finite or infinite) collection 
of interacting subpopulations, each consisting of a (finite or infinite)
number of identical units;
sizes of the subpopulations and couplings between them are generally different
(cf. \cite{Bareto-Hunt-Ott-So-08,Ott-Antonsen-08} and references therein).
Using the seminal approach of  Watanabe and Strogatz
(WS)~\cite{Watanabe-Strogatz-94},
we demonstrate that each subpopulation can be described by only three dynamical 
variables plus constants of motion, determined by initial conditions.
This partial integrability allows us to separate the full dynamics
 into a relatively
small number of generally dissipative modes 
(their number is proportional to the number of subpopulations) with
possibly nontrivial behavior, and the integrals of motion.
In particular, our theory allows us to extend some recent 
results \cite{Abrams-Mirollo-Strogatz-Wiley-08,Ott-Antonsen-08} 
and to determine the conditions of their validity.  

Our basic model is a generalization of the Kuramoto 
model~\cite{Kuramoto-84}, cf. \cite{Bareto-Hunt-Ott-So-08,Ott-Antonsen-08}:
\be
\frac{d\phi^a_k}{dt}=\w_a+\frac{1}{N}\sum_{b=1}^{M}\sum_{j=1}^{N_b}
\e_{a,b}\sin (\phi^b_j-\phi^a_k-\alpha_{a,b})\;.
\ee{eq1}
Here we denote the subpopulations by indices $a,b=1,\ldots,M$. Variable
$\phi^a_k(t)$ is the phase of 
oscillator $k$ in subpopulation $a$; $k=1,\ldots N_a$, where $N_a$ is the 
size of the subpopulation, and
$\w_a$ is the natural frequency of its oscillators (we remind that all
oscillators in a subpopulation are {\em identical}).
The total number of oscillators is $N=\sum N_a$ and two constants $\e,\alpha$ 
describe the coupling with an arbitrary phase shift, cf.~\cite{Sakaguchi-Kuramoto-86}.
The system can be re-written as
\begin{align}
\frac{d\phi^a_{k}}{dt}&=\w_a+\text{Im}(Z_ae^{-i\phi_k^a})\;,
\label{eq2}\\
Z_a&=\sum_b n_b \e_{a,b}e^{-i\alpha_{a,b}}r_b e^{i\Theta_b}\;,
\label{eq3}
\end{align}
where $Z_a$ is the effective force acting on the oscillators of subpopulation $a$.
Here we have introduced the relative population sizes $n_a=N_a/N$ and 
the complex mean fields for each subpopulation
\be
X_a+iY_a=r_a e^{i\Theta_a}=
N_a^{-1}\sum_{k=1}^{N_a}e^{i\phi^a_{k}}\;.
\ee{eq4}
Note that all oscillators in a subpopulation obey the same equation,
though generally they
have different initial conditions $\phi^a_{k}(0)$. 
Thus, we can apply to \textit{each subpopulation} 
the WS ansatz \cite{Watanabe-Strogatz-94}
that reduces the dynamics of the subpopulation to that of three 
variables $\rho_a(t)$, $\Psi_a(t)$, $\Phi_a(t)$, via the transformation
\footnote{For convenience we use variables, different from those in 
\cite{Watanabe-Strogatz-94}.}
\be
\tan\left[\frac{\phi^a_{k}-\Phi_a}{2}\right]=\frac{1-\rho_a}{1+\rho_a}
\tan\left[\frac{\psi^a_{k}-\Psi_a}{2}\right]\;
\ee{eq5}
containing $N_a$ constants $\psi^a_{k}$, which are directly determined
from the initial state $\phi^a_{k}(0)$
and additionally satisfy 
\be
\sum_{k=1}^{N_a}\cos\psi^a_{k}=\sum_{k=1}^{N_a}\sin\psi^a_{k}=0\;.
\ee{eq6}
Due to an arbitrary shift of constants $\psi_k$ with respect to $\Psi$, 
only $N_a-3$ of constants $\psi^a_k$ are independent.
The WS method is valid generally, provided the number of oscillators in a 
subpopulation is larger than three, and the initial state does 
not have too many clusters, see~\cite{Watanabe-Strogatz-94} for 
a detailed discussion of these conditions and of how 
$\rho_a(0)$, $\Psi_a(0)$, $\Phi_a(0)$, and $\psi^a_{k}$ can be computed from
$\phi^a_{k}(0)$.
With account of Eq.~(\ref{eq3}), we write the WS equations for our setup as
\begin{align}
\frac{d\rho_a}{dt}&=\frac{1-\rho_a^2}{2}\mbox{Re}(Z_ae^{-i\Phi_a})\;,          \label{eq7-1}\\
\frac{d\Psi_a}{dt}&=\frac{1-\rho_a^2}{2\rho_a}\mbox{Im}(Z_ae^{-i\Phi_a})\;,      \label{eq7-2}\\
\frac{d\Phi_a}{dt}&=\w_a+\frac{1+\rho_a^2}{2\rho_a}\mbox{Im}(Z_ae^{-i\Phi_a}) \;. \label{eq7-3}
\end{align}

In order to illustrate the physical meaning of the new variables, 
let us consider how they characterize
the distribution of the phases of a subpopulation. Generally,
oscillators form a bunch, and the amplitude 
$\rho$ characterizes its width: 
$\rho=0$, if the distribution is uniform (asynchrony) and $\rho=1$, 
if the distribution shrinks to $\delta$-function (full synchrony).
Amplitude $\rho$ is roughly proportional to the amplitude of the
mean field $r$ (see Eq.~\reff{eq4}) in the sense that $\rho=r=0$ for the
full asynchrony  and $\rho=r=1$ for the full synchrony.
For intermediate cases these quantities generally differ and coincide only in 
a special case, outlined below.
The phase variable $\Phi$ characterizes the position of the bunch, 
and is therefore related to the phase of the mean field, 
$\Phi\approx\Theta$.
Another phase variable $\Psi$ describes 
the motion of individual oscillators with respect to the bunch
(generally, the oscillators can move with a velocity different from that
of the bunch, see
\cite{Rosenblum-Pikovsky-07} for an example of such a dynamics). 

The set of Eqs.~(\ref{eq7-1}-\ref{eq7-3}) is a straightforward generalization 
of the WS equations \cite{Watanabe-Strogatz-94} to the case of $M$ 
interacting subpopulations. 
For a further analysis, and in particular for the consideration of the
thermodynamic limit, it is convenient to introduce new variables, 
a phase shift $\zeta_a=\Phi_a-\Psi_a$ 
and a complex bunch amplitude
$z_a=\rho_a e^{i\Phi_a}$.
Then 
we can rewrite Eqs.~(\ref{eq7-1}-\ref{eq7-3}) as
\begin{align}
\frac{dz_a}{dt}&=i\w_a z_a+\frac{1}{2}Z_a-\frac{z^2_a}{2}Z^*_a\;,\label{eq9-1}\\
\frac{d\zeta_a}{dt}&=\w_a+\text{Im}(z_a^*Z_a)\;.\label{eq9-2}
\end{align}
Next, we have to represent the complex force $Z_a$ (see Eq.~\reff{eq3}) in
terms of new variables. For this goal it is convenient to rewrite Eq.~\reff{eq5}
in an equivalent form 
$e^{i\phi_k}=e^{i\Phi}(\rho e^{i\Psi}+ e^{i\psi_k})/(\rho e^{i\psi_k}+e^{i\Psi})$.
Substituting this into Eq.~\reff{eq4}, we obtain:
\begin{equation}
\begin{aligned}
r_a e^{i\Theta_a}&=\rho_a e^{i\Phi_a}\gamma_a(z_a,\zeta_a)=z_a \gamma_a(z_a,\zeta_a)\;,\\
\gamma_a(z_a,\zeta_a)&  
=\frac{1}{N_a}\sum_{k=1}^{N_a}\frac{1+|z_a|^{-2}z^*_a
e^{i(\zeta_a+\psi^a_{k})}}{1+z^*_a e^{i(\zeta_a+\psi^a_{k})}}\;.
\end{aligned}
\label{eq10-3}
\end{equation}

From Eq.~(\ref{eq9-1}) it follows that the dynamics of the complex bunch
amplitude 
of a subpopulation $z_a$ is determined by the force $Z_a$, resulting from interaction 
within the subpopulation as well as from interaction with other subpopulations.
Contributions to $Z_a$ are proportional  to the relative weights $n_b$,
to the coupling constant
$\e e^{-i\alpha}$ 
 and to the complex mean field
$\gamma_b(z_b,\zeta_b) z_b$, which generally depends not only on the global 
variables $\zeta_b$ and $z_b$, but also on the constants of motion $\psi^b_{k}$.
Equations (\ref{eq9-1}-\ref{eq9-2}), as well as equivalent equations
(\ref{eq7-1}-\ref{eq7-3}), together with the definitions
(\ref{eq3},\ref{eq10-3}) are exact and complete; they show that the dynamics of a 
hierarchical ensemble of oscillators can be reduced to $3M$ ODEs 
plus $N-3M$ constants of motion. Before proceeding with the analysis of these 
equations and examples, let us discuss how a thermodynamic limit
$N\to\infty$ can be introduced in this picture. There are two main
ways of performing this. 

(i) Suppose that the number of subpopulations $M$ 
remains finite, but their sizes grow $N,N_a\to\infty$ in a way that $n_a=const$.
In this case only Eq.~\reff{eq10-3}  is affected and should be now written as an integral
\be
\gamma_a(z_a,\zeta_a)=\int_{-\pi}^{\pi}
\frac{1+|z_a|^{-2}z^*_a
e^{i(\zeta_a+\psi)}}{1+z^*_a e^{i(\zeta_a+\psi)}}\sigma_a(\psi)\;d\psi\;.
\ee{eq11}
Here $\sigma_a(\psi)$ is the distribution of the constants of motion $\psi$ in the 
subpopulation $a$, additionally it satisfies (cf. \reff{eq6})
\be
\int_{-\pi}^{\pi}\sigma_a(\psi)e^{i\psi}\;d\psi=0\;.
\ee{eq12}
In this limit the ensemble is described by a set of $3M$ ODEs, where the right 
hand sides depend on the variables via integrals \reff{eq11}.
The integrals of motion are now the functions $\sigma_a(\psi)$.

(ii) In another limiting case, we keep the size of each subpopulation
$N_a$ finite but let the number of subpopulations grow $M\to\infty$. 
Considering indices $a,b$ 
as continuous variables, we write instead of Eq.~\reff{eq3}
\be
Z(a)=\int db\; n(b) [\e(a,b)+i\eta(a,b)] \gamma(b) z(b)\;.
\ee{eq13}
Now Eqs.~(\ref{eq9-1}-\ref{eq10-3},\ref{eq13}) become a system of 
integral equations; still it is simpler than the original equation \reff{eq1} 
as at each value of the continuous parameter $a$ we have only three real 
time-dependent variables.

Certainly, one can also perform both thermodynamic limits
simultaneously. Then the ensemble is described by the system
(\ref{eq9-1},\ref{eq9-2},\ref{eq11},\ref{eq13}).

Next we study 
an important case when Eqs.~(\ref{eq9-1},\ref{eq9-2})  \textit{decouple}.
To this end we represent the fraction in Eqs.~(\ref{eq10-3},\ref{eq11}) as a series 
\be
\gamma_a=1+\left(1-|z_a|^{-2}\right)\sum_{l=2}^{\infty}C^a_l\left(-z_a^*e^{i\zeta_a}\right)^l \;,
\ee{eq14}
where complex constants $C^a_l$ depend only on the distribution of the
constants of motion
\be
C^a_l=\frac{1}{N_a}\sum_{k=1}^{N_a}e^{il\psi^a_k}
\;\;  \mbox{or}\;\;
C^a_l=\int_{-\pi}^{\pi} \sigma_a(\psi)e^{il\psi}\;d\psi\;,
\ee{eq15}
and we used that $C^a_{1}=0$ due to Eqs.~(\ref{eq6},\ref{eq12}).
Obviously, the governing equations simplify, if
$C^a_l=0$ for $l\geq 2$ and all $a$, and, hence, $\gamma=1$.
Then the force $Z$ does not depend on the phase
variable $\zeta$ and 
Eq.~(\ref{eq9-1}) decouples from Eq.~(\ref{eq9-2}).  
It is easy to see from Eqs.~\reff{eq15} that 
$C^a_l$, which are in fact Fourier coefficients of the distribution of the
constants of motion $\psi$, vanish in the thermodynamic
limit of type (i), if $\sigma(\psi)=1/2\pi$. However, 
if the number of oscillators in a subpopulation $N_a$
is finite, then even for a uniform spreading of $\psi_k$, 
the discrete sum in \reff{eq15}
yields $|C^a_l|=1$, $\mbox{arg}(C^a_l)=\psi^a_1$ for $l=N_a,2N_a,\ldots$, and we get 
\be
\gamma_a=1+\frac{1-|z_a|^2}{1-\left [-z_a^*e^{i(\zeta_a+\psi_1^a)}\right ]^{N_a}}
\left [ -z_a^*e^{i(\zeta_a+\psi_1^a)}\right ]^{N_a}\;.
\ee{eq175}
Thus, the deviation of $\gamma_a$ from unity is exponentially small in the size of 
the subpopulation and, therefore, can be neglected for large $N_a$.
This is exactly the case, where the complex bunch amplitude $\rho$ is equal to
the mean field amplitude $r$, because in \reff{eq10-3} $\gamma=1$.

Hence, for the uniform distribution of constants of motion $\psi$, 
ensemble \reff{eq1}
admits a simplified description via Eq.~\reff{eq9-1}, supplemented by an equation for $Z_a$, 
either in a discrete or in a continuous (for the thermodynamic limit of
type (ii)) form: 
\begin{align}
Z_a&=\sum_b n_b \e_{a,b}e^{-i\alpha_{a,b}}  z_b \label{eq17-2} \;,\\ 
Z(a)&=\int db\; n(b) \e(a,b)^{-i\alpha(a,b)}  z(b)\;.\label{eq17-3}
\end{align}
A relation between the distribution of the
original phases $\phi_k$ and the uniform distribution of constants of
motion $\psi_k$ follows from Eq.~\reff{eq5}: one can see that
different distributions of the phases $\phi_k$, parameterized by
different values of $\rho$, correspond to the uniformly distributed
constants $\psi_k$. 

As a first application of our framework, we apply 
Eqs.~(\ref{eq9-1},\ref{eq9-2},\ref{eq10-3},\ref{eq13}) 
to the classical Kuramoto problem (cf.~\cite{Ott-Antonsen-08}).  
We set $\e(a,b)=\e=\mbox{const}$, $\alpha=0$, use the frequency as the subpopulation index, 
$a=\w$, and perform
the thermodynamic limit (ii). As a result, in the case when $\gamma=1$
and the variable $\zeta$ (as well as the constants of motion) 
does not influence the dynamics,  we obtain exactly Eqs.~(\ref{eq9-1},\ref{eq17-3}), derived
recently by 
Ott and Antonsen (OA)~\cite{Ott-Antonsen-08} under an assumption of a
certain parameterization of the phase distribution. 
Considering the Lorentzian distribution of natural frequencies
$n(\omega)=[\pi(\omega^2+1)]^{-1}$ and using analytic properties of $z(\omega)$ 
as a function of complex
frequency $\omega$, OA have calculated the integral in Eq.~\reff{eq17-3} 
by the residue of the pole at $\omega=i$ and have obtained $Z=\e z(i)$. 
Substituting $Z$ into Eq.~\reff{eq9-1} for $\omega=i$, OA derived a
closed equation for $\tilde Z=Z/\e$, i.e. for the usual Kuramoto mean field of the 
whole population:
\be
\dot {\tilde Z}=(-1+\frac{\e}{2})\tilde Z-\frac{\e}{2}|\tilde Z|^2\tilde Z\;, 
\ee{eq18}
solved it, and in this way obtained explicitely
the evolution of the mean field.

From our derivation of the equations of motion we conclude, 
that the particular ansatz used in 
\cite{Ott-Antonsen-08}
corresponds to the case of uniformly distributed constants of motion $\psi_k$, 
what is equivalent to 
vanishing Fourier coefficients $C_l$. Next we discuss, what changes if the
distribution of constants $\psi_k$ is not uniform, i.e. $C_l\ne 0$.
Let us treat the effect of non-vanishing coefficients $C_l$ 
perturbatively, assuming that in the first approximation the 
OA ansatz is valid. 
Considering for simplicity the effect of $C_2\ne 0$ only, 
we obtain a correction to the mean field by 
substituting \reff{eq14} into \reff{eq13}
\be
\Delta Z\approx \e\int \;d\omega
\frac{z^*(\omega)(|z(\omega)|^2-1)e^{i2\zeta(\omega)}C_2(\omega)}{\pi(\omega^2+1)}
\;.
\ee{eq19}
Calculation of this integral by the residue yields
\be
\Delta Z\approx\e z^*(i)(|z(i)|^2-1)e^{i2\zeta(i)}C_2(i)\;.
\ee{eq20}
From Eq.~\reff{eq9-2} it follows that in the first approximation $\zeta(i)=\zeta_0+it$. 
Therefore $\Delta Z\propto e^{-2t}$. We conclude that the  contribution from a nonuniform 
distribution of constants $\psi_k$ results in an exponentially decaying correction to the mean field. 
The characteristic time scale of this decay is $1/2$, to be compared with the characteristic time scale 
of the evolution of the mean field, which, according to \reff{eq18}, is $(\e/2-1)^{-1}$. 
Thus, close to criticality $\e_c=2$, the approximation of vanishing constants $C_l$ works 
well after 
short transients; this is not surprising as near a bifurcation point the
dynamics is typically effectively low-dimensional, dominated by a few normal modes. 
Far from criticality the time scale separation is not valid and the
dynamics is generally high-dimensional.

\begin{figure}[!tbh]
\centerline{\includegraphics[width=0.42\textwidth]{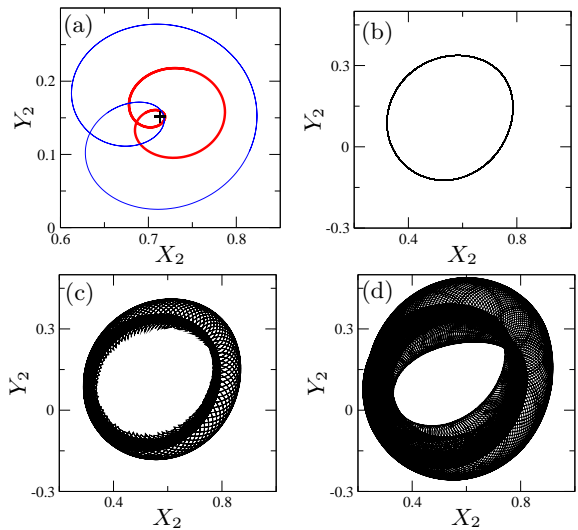}}
\caption{(Color online) 
Simulation of ensemble \reff{eq1} for $N_1=N_2=64$, $\alpha=\pi/2-0.1$,
and different distributions of constants of motion $\psi_k^{(2)}$. 
Mean fields $X_2$, $Y_2$ are defined by Eq.~\reff{eq4}.
(a): $\mu=0.6$; uniform distribution of $\psi_k^{(2)}$
results in the steady state (black plus), cf.~\cite{Abrams-Mirollo-Strogatz-Wiley-08}, 
whereas nonuniform distributions with $q=0.9$ and $q=0.7$ yield limit cycle solutions 
(bold red and blue solid lines, respectively).
(b)-(d): $\mu=0.65$; uniform distribution of $\psi_k^{(2)}$ yields a limit cycle 
solution (b),
cf.~\cite{Abrams-Mirollo-Strogatz-Wiley-08}, whereas  for $q=0.9$ (c) and for 
$q=0.7$ (d) we observe a new type of the chimera state with a quasiperiodic dynamics.
}
\label{fig1}
\end{figure}

As a second example we extend recent results 
of Abrams \textit{et al}~\cite{Abrams-Mirollo-Strogatz-Wiley-08}. 
They studied two coupled subpopulations of identical oscillators, i.e. 
model~\reff{eq1} with $a=1,2$, $\omega_1=\omega_2$, $N_1=N_2$ and 
heterogeneous coupling
$\e_{1,1}=\e_{2,1}=2\mu$, 
$\e_{1,2}=\e_{2,1}=2\nu$, and $\alpha_{a,b}=\alpha$, where $\nu=1-\mu$.
Using the OA ansatz~\cite{Ott-Antonsen-08}, Abrams \textit{et al} derived equations for the 
complex order parameters $z_{1,2}$ and analyzed the so-called chimera state, where, e.g., the first 
subpopulation is fully synchronized, $\rho_1=1$, whereas the other one is only partially
synchronized, $\rho_2<1$;
they have found both static and time-periodic solutions for $\rho_2$.
With our approach we describe the system {\em exactly}, by writing six 
Eqs.~(\ref{eq7-1}-\ref{eq7-3}) for both subpopulations.
Since we are interested in the chimera state in the second subpopulation, 
the first, synchronous one, is described by its phase $\Phi_1$ only. 
In this case  
$Z_1=\mu e^{i(\Phi_1-\alpha)}+\nu A e^{i(\Phi_2+\beta-\alpha)}$,
$Z_2=\nu e^{i(\Phi_1-\alpha)}+\mu A e^{i(\Phi_2+\beta-\alpha)}$,
where 
\be
A(\rho_2,\Psi_2)e^{i\beta(\rho_2,\Psi_2)}=\frac{1}{N_2}\sum_{k=1}^{N_2}
\frac{\rho_2e^{i\Psi_2}+e^{i\psi_k^{(2)}}}{e^{i\Psi_2}+\rho_2e^{i\psi_k^{(2)}}}\;.
\ee{eq22}
Next, we note that the dynamics depends only on the phase difference 
$\delta=\Phi_1-\Phi_2$ and, hence, write a closed system of three equations:
\begin{align}
\frac{d\rho_2}{dt}&=\frac{1-\rho_2^2}{2}
(\mu A\cos(\beta-\alpha)+\nu\cos(\delta-\alpha))\;,   \label{eq21-1}\\
\frac{d\delta}{dt}&=-\mu(\sin\alpha+\frac{1+\rho_2^2}
{2\rho_2}A\sin(\beta-\alpha))\nonumber\\
&+\nu(A\sin(\beta-\alpha-\delta)-\frac{1+\rho_2^2}{2\rho_2}\sin(\delta-\alpha))\;, \label{eq21-2}\\
\frac{d\Psi_2}{dt}&=\frac{1-\rho_2^2}{2\rho_2}[\mu
A\sin(\beta-\alpha)+\nu\sin(\delta-\alpha)]\;.  \label{eq21-3}
\end{align}
Following Abrams \textit{et al}~\cite{Abrams-Mirollo-Strogatz-Wiley-08}
we take a thermodynamic limit $N_2\to\infty$. 
Next, we take a uniform distribution of the constants
$\psi^{(2)}$ (which enter only  via the relations
\reff{eq22}) -- we remind that this choice corresponds to the restriction 
imposed by OA on the phase distribution in their ansatz. 
Then from Eq.~\reff{eq22} it follows $A=\rho_2=r_2$, $\beta=0$
and equations for $\rho_2$ and $\delta$ decouple from Eq.~\reff{eq21-3}.
The obtained Eqs.~(\ref{eq21-1},\ref{eq21-2}) constitute exactly the system 
analyzed in \cite{Abrams-Mirollo-Strogatz-Wiley-08}.
For a nonuniform distribution of $\psi_k^{(2)}$, we
have to analyze the full three-dimensional system (\ref{eq21-1}-\ref{eq21-3}), 
which certainly can exhibit more complex solutions.

To verify our theoretical prediction, we have performed numerical simulations 
of the ensemble \reff{eq1} for the same parameters, where 
Abrams \textit{et al} obtained stationary and time-periodic solutions, 
but for different distributions of the constants $\psi_k$.
Namely, we took $\psi_k$, uniformly distributed in the range
$-q\pi<\psi_k^{(2)}<q\pi$, where $q\le 1$ is a parameter.
For $q=1$ we have reproduced the results of 
\cite{Abrams-Mirollo-Strogatz-Wiley-08}, 
while for $q<1$ the dynamics attains an additional time dependence and
becomes periodic and quasiperiodic, respectively (see Fig.~\ref{fig1}).

In conclusion, we have performed the exact reduction of the dynamics of 
hierarchically organized populations of coupled oscillators. Due to the
partial integrability, only three dynamical variables remain relevant
for each subpopulation, all other are constants of motion. This reduced
description is independent of the subpopulation sizes and
holds also in the thermodynamic limit. In an important particular case,
when the distribution of the constants of motion is uniform, the
governing equations further decouple and simplify. We have demonstrated
that this case corresponds to the recently found particular ansatz of
Ott and Antonsen~\cite{Ott-Antonsen-08}. The analysis of full equations
has allowed us to estimate corrections to this particular solution due
to non-uniformity of constants. Application of our framework to the 
model by Abrams \textit{et al} \cite{Abrams-Mirollo-Strogatz-Wiley-08} revealed
existence of novel, quasiperiodically breathing chimera states.

In this Letter we restricted our attention to the simplest setups in order 
to demonstrate applicability of the theory. However, the method can be in a 
straightforward way extended to the cases of
nonlinearly coupled populations~\cite{Rosenblum-Pikovsky-07}, externally forced 
ensembles, etc. In these cases even a chaotic dynamics of the global
variables can be expected.
The main limitation of the theory is that the coupling in
Eq.~\reff{eq1} has a sine form.

We acknowledge financial support from DFG (SFB 555).


\end{document}